# Coverage Related Issues in Networks

**Marida Dossena*[1]**

[1]Department of Information Sciences, University of Naples Federico II, Napoli, Italy
Email: marida.dossena@libero.it

**Abstract**- Wireless sensor networks consisting of great number of cheap and tiny sensor nodes which are used for military environment controlling, natural events recording, traffic monitoring, robot navigation, and etc. Such a networks encounter with various types of challenges like energy consumption, routing, coverage, reliability. The most significant types of these problems are coverage that originated from the nodes energy consumption constrained. In order to dominate this problem different kinds of methods has been presented where the majority of them based on theoretical methods and used unbalanced and calculated distributions. In all of the proposed methods a large numbers of nodes are used. In this paper our attempt is based on using a few numbers of sensors in order to cover the vast area of environment. We proposed an algorithm that divides the desired environment to several areas and in each of these areas by using the genetic algorithm improve the coverage. The proposed method is simulated in MATLAB software and the obtained results are compared with the existing algorithms. Results show that the presented algorithm has a substantial coverage in compare with its previous counterparts.

___________________________________________________________________

*Corresponding author. E-mail: marida.dossena@libero.it

# 1- INTRODUCTION

wireless sensor networks are high level distributed network consisting of large number of light, cheap and tiny sensor nodes which are used for environmental and complex systems monitoring. Each node includes three subsystems. First subsystem of sensor node, sense the environment, second one is processing unit which do the calculating operations and the third part is communication subsystem that is in charge of transmitting data among neighbor nodes [1]. Sensor nodes have restrictions in processing and sensing abilities. In wireless sensor networks node's placement is not predetermined. So this criterion enables us to spread them in dangerous and unavailable environments [2]. In recent years majority of attempts are based on distributing sensor nodes in applicable environments. Power constraint in wireless sensor networks relates to the main restrictions such as power consumption, coverage and network lifetime [3]. In order to achieve accurate monitoring, the better coverage should be exist. In designing a network various kinds of node distribution strategies which relate to the heterogeneous coverage and energy consumption and also some the other criterions must be used [3]. One of the effective methods in node spreading to have the better power consumption and coverage is control capability of sensor nodes location. But we have to demonstrate that, mentioned methods are relatively practical in some conditions that the numbers of sensor nodes are great. Regional position of network maybe unavailable physically for programmer due to geographic restrictions. In such cases, sensor nodes spread non - uniform and randomly in environment [4]. In paper [5] where sensor distribution is random, coverage probability is obtained as follow:

$$f = 1 - e \quad (1)$$

Where rs is sensing range and λ is nodes density. By increasing these parameters, coverage increase either. But in cited method if the nodes density become high in the last hop toward the base station, nodes around sink die earlier than other sensor nodes. Another method is using nodes with more energy and lifetime in compare with other nodes around base station. In a time of pointing to some uniform distribution methods, Gaussian protocol attracts attention. In this method base station stay in the middle of the environment and by using Gaussian distribution sensor nodes spread in there [3]. Although Gaussian protocol can increase lifetime and coverage but equally it can increase cost. Our attempt in this paper is based on using a few numbers of sensors to achieve great amount of coverage and yet decrease the cost. In recent years different kinds of algorithms for coverage protection in wireless sensor networks are suggested where majority of them use node redundancy to achieve this issue. Some times when there is no demand for node's sense, mentioned algorithms keep them in sleep mode [6]. Later, algorithms based on search operation were suggested that control the node's sleep mode to achieve more coverage in network [7]. In this paper we propose a method on a base of genetic algorithm which can decrease education time. In mentioned method environment is divided to several subsystems. We try to cover desired area with few numbers of nodes and less amount of energy consumption. Although previous algorithms have better coverage but they use large number of sensor nodes which relate to the cost increasing.

The rest of the paper is organized as follow: section 2 presents existing related works, section 3 describes the proposed algorithm, section 4 evaluates the proposed algorithm and compare it with the existing counterparts and section 5 concludes the paper.

## 2. RELATED WORKS

Some of the existing algorithms for coverage and energy consumption spread nodes non - uniform in wireless sensor network. In paper [8] sensor nodes are distributed like mathematical models. When sensor nodes spread non - uniform in the environment, in one - hop toward base station due to energy constraint, energy hole will be happen and this is one of the non - uniform spreading problems [9, 10]. In paper [4] avoiding energy hole by distributing heterogeneous sensors and using different data reporting time has been discussed. Energy model will be as follow:

$$E = d^{\alpha} + c$$

(2)

D is transmission distance, α > 2 and C is maximum transmission range. When α > 2 energy will decrease the life time of such networks so much and relates to the increasing sensor redundancy near the base station. Sometimes sensors with high amount of energy around base station are used. In some networks that their sensors spread non - uniform and homogeneous, when the initial energy of nodes remains more than %90, network life time finished [10 - 14]. For dominating this problem Gaussian protocol is presented in a way that base station placed in the middle of the environment and by using two dimensional Gaussian distribution, spread sensor nodes in there. Figure 1 demonstrates two dimensional Gaussian distribution [3].

Two dimensional Gaussian distribution is formulated as follow:

$$(3)$$

(x,y) refers to the base station location coordinates and base station place in (0,0), equation 4 is rewritten as follow:

$$(4)$$

$A_x$, $B_y$ is standard deviation for x,y measures [3]. In paper [15] distributed self spreading method with uniform distribution for optimizing coverage is presented. In mentioned algorithm, coverage is defined by means of covered areas connection by each node to the whole space of sensing environment and nodes uniform distribution as an average of local inner- nodes distance standard deviation. In such kinds of distributions, due to uniform distance among nodes energy consumption is balance. Nodes distribute randomly and each node is aware of its own location. This method use electrical power that depends on a local density µ and the difference of distance among nodes. At first density of each node is equal to the number of neighbor nodes. Electrical power into the i'th node by j'th node in n'th hop is calculated as equation (5).

$$\text{(5)}$$

Where the $P_{i}^{n}$ is the location of i'th node in the n'th hop. The next movement of a node depends on the power imported from neighbor node to it. The algorithm will be stop in a time that a node move for infinite time in a short timeframe or come back to the previous location and just move between two fix location. In paper [16] an algorithm is presented that use portable and static sensors simultaneously and by means of auction, increase coverage. At first portable and static nodes spread randomly in the environment. Then static sensors determine their Voronoi polygon and find the coverage area and the nearest portable sensors to desired hole will transmit data. Voronoi polygon is shown in figure 2[17].

The farthest Voronoi vertex has been selected as a portable sensor's destination and auction rate is allocated to each of the portable nodes. Initial value of auction rate is 0. Auction rate has been calculated as follow:

$$\text{Ratio Bidding} = \pi (d - R_s)^2 \qquad (6)$$

D is a distance between sensor and the farthest Voronoi vertex, Rs is the sensing radius. A static sensor send auction message to the nearest portable sensor that its main rate is lower than the suggested rate. Portable sensor receives all of the auction messages from neighbors and selects the highest suggestion and move to fill its coverage hole. The highest auction rate is recorded as a new rate for portable sensor. This method guarantees that portable sensor first cover the biggest hole.

3. PROPOSED ALGORITHM

Genetic algorithm is one of the heuristic methods for solving optimization problems and is on a base of random search and iteration. In proposed algorithm if number of nodes is large and environment is vast, it will be divided to subareas and execute genetic algorithm in each subarea in a parallel manner to decrease the education time. Dividing area based on the number of nodes and area space is one of genetic algorithm deficiency. In this method at first nodes spread randomly and then by using genetic algorithm nodes are educated. The maximum range that a sensor can cover is like a circle with Rs radius which is shown in figure 3. In order to obtain coverage of the whole node, coverage of each node is calculated separately and divided to the desired area space. There is possibility in such methods that nodes have overlapping with each other or physical obstacles exist which relates to the coverage decreasing. Coverage can be measured as equation7:

$$\qquad\qquad (7)$$

Ci is each node's coverage space and RS2 is a desired space. First nodes are spread in the whole environment randomly and then the environment is divided to subarea based on a number of nodes and environment size. Nodes are spread in the environment for 100 times and this number of node's spreading is selected as a first population. Existing chromosomes in the algorithm are existing nodes in the subarea and genes identified a sensor node along with its location coordinates and coverage space and its own ID. Chromosomes' coding is shown in figure 4 that consisting of n nodes or genes.

Fitness function of each chromosome is calculated and from the sum of all chromosomes or subarea, the whole coverage percent is obtained. Fitness function is as follow and its parameters are described in table 1.

$$\qquad\qquad (8)$$

Table 1. Fitness function parameters

| parameters | description |
|---|---|
| p | Number of subarea |
| $GC_i$ | Coverage of nodes |
| n | Number of subarea nodes |
| $R^2$ | area |

After doing all of aforementioned steps, average population will be generated. These operations are done in a way that the special operation of genetic algorithm which is cutting and jumping is done on each subarea. In cutting operation, two standard points and jumping are used and in order to avoid environment turbulence and not losing parents' characteristics, %1 of the first population is selected as a new population. Cutting and jumping rate are determined 0.05 and 0.85 respectively. When cutting operation is done, three chromosomes are selected randomly and by using Roulette wheel two of them are selected and cutting is done. Two chromosomes will be added to the middle population and for jumping, two chromosomes will be selected and jumping will be done. Due to this fact that achieved middle population is twice first population, fitness function of each chromosome is calculated and is ordered on a base of fitness function. According to the numbers of first population, best chromosomes are selected and these operations up to a time of reaching stop condition will be repeated. Stop condition can be constant number or be the difference between coverage percent of two generation if it is lower than 0.001. In this paper, two generation coverage difference is used. In figure 5 pseudo code of the proposed algorithm is demonstrated.

The cited algorithm is repeated for all of the nodes and in the node number 586, coverage is %99.9, which is shown in Figure 6. Each of the nodes is repeated in 50 generation, then average repetition rate are determined as a coverage level of nodes. Where in figures 7, 8, 9 for different number of nodes are shown and in table 2, coverage percent of each node is demonstrated briefly.

Tabel2: Coverage of proposed algorithm

| Number of node | proposed algorithm |
|---|---|
| 50 | 30.9371 |
| 100 | 43.7516 |
| 150 | 50.5200 |
| 200 | 59.5384 |
| 250 | 66.4327 |
| 300 | 72.9193 |
| 400 | 84.6200 |
| 500 | 93.95 |
| 550 | 989833 |
| 586 | 99.99 |

4. COMPARING THE PERFORMANCE OF PROPOSED ALGORITHM WITH PREVIOUS COUNTERPARTS

**Proposed algorithm is simulated in MATLAB software and obtained results are compared with two dimensional Gaussian protocols in figure 10. Two dimensional Gaussian protocol achieve %99.9 coverage in 1000 nodes, however in the proposed algorithm, mentioned coverage percent has been achieved just in 580 numbers of nodes. Therefore genetic algorithm in compare with two dimensional Gaussian protocol cover more area with less number of nodes. In the proposed algorithm if the number of nodes exceeds 580, algorithm performance will decrease. If extra number of nodes added to it, it will just increase overlapping and won't optimize coverage.**

Table 3. Comparing coverage percent of proposed algorithm with existing methods

| Number of node | Gaussian protocol | proposed algorithm |
|---|---|---|
| 100 | 35 | 43.7516 |
| 200 | 51 | 59.5384 |
| 300 | 63 | 72.9193 |
| 400 | 74 | 84.6200 |
| 500 | 82 | 93.95 |
| 600 | 88 | 99.99 |
| 700 | 92 | 99.99 |
| 800 | 95 | 99.99 |
| 900 | 97 | 99.99 |
| 1000 | 99.99 | 99.99 |

5. CONCLUSION

**In this paper a method based on genetic algorithm which can cover the certain range of area with less number of nodes in compare with the previous algorithms is proposed. In the presented algorithm the number of sensor nodes decreased nearly half of the number of nodes that used in recent algorithms. Also among the nodes which have great amount of overlapping near base station, just one of them**

send information and the following operation relates to not creating energy hole near base station. In algorithms which their nodes spread non – uniform when network lifetime finished, %90 of node's first energy remains. This problem can be solved by presenting weight to each of the service quality by using weighted genetic algorithm and optimize the desired services simultaneously.

Figure 1:

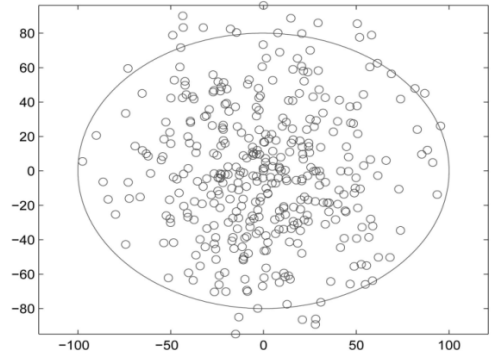

**Figure 2:**

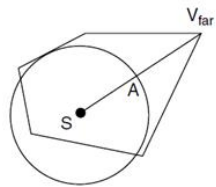

**Figure 3:**

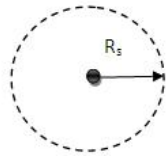

**figure 4:**

| node$_1$ | node$_2$ | node$_3$ | node$_4$ | … | node$_n$ |
|---|---|---|---|---|---|
| ↓ | ↓ | ↓ | ↓ | | ↓ |
| ID$_1$ | ID$_2$ | ID$_3$ | ID$_4$ | … | ID$_n$ |
| X$_1$ | X$_2$ | X$_3$ | X$_4$ | … | X$_n$ |
| Y$_1$ | Y$_2$ | Y$_3$ | Y$_4$ | … | Y$_n$ |
| C$_1$ | C$_2$ | C$_3$ | C$_4$ | … | C$_n$ |

**Figure 5:**

```
Initial population size
Random _ population;
subdivide  area sensing in equal part;
ratio crossover=0.85;
ratio mutation=0.05;
while( subtraction (Fitness(i) ,Fitness( i+1))<0.001)
for i=2: population size
     selection(old_ population);
     selection(old_ population);
     selection(old_ population);
     r=rand;
     if(r< ratio crossover)
selection( Roulette wheel );
for i=1: 6
        crossover two chromosome
     end//for
end//if
     r=rand;
     if(r< ratio mutation)
        mutation( old_ population );
     end//if
     r=rand;
     if(r< ratio mutation)
mutation( old_ population );
     end//if
     next _ population ();
fitness( best(next _ population) );
cut _ population
end//while
```

**Figure 6:**

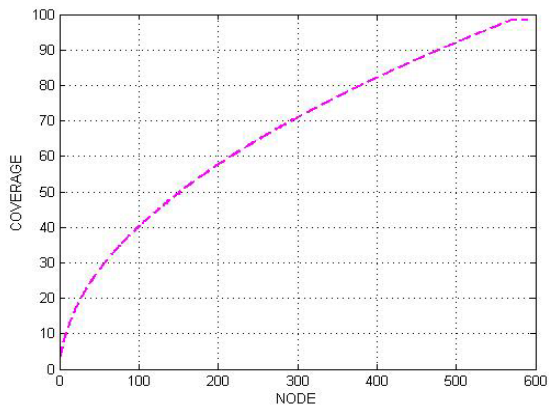

Figure 7:

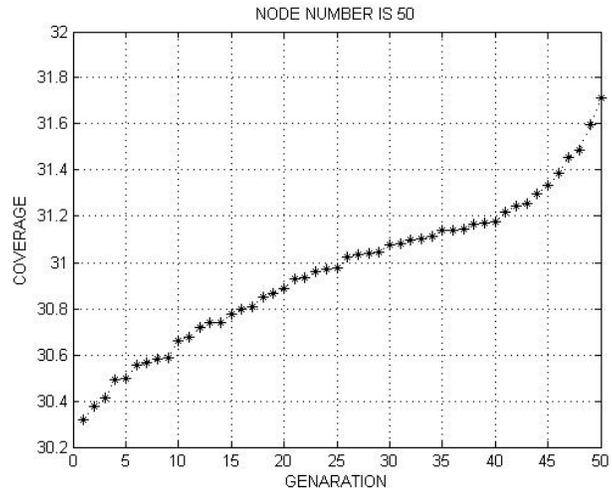

Figure 8:

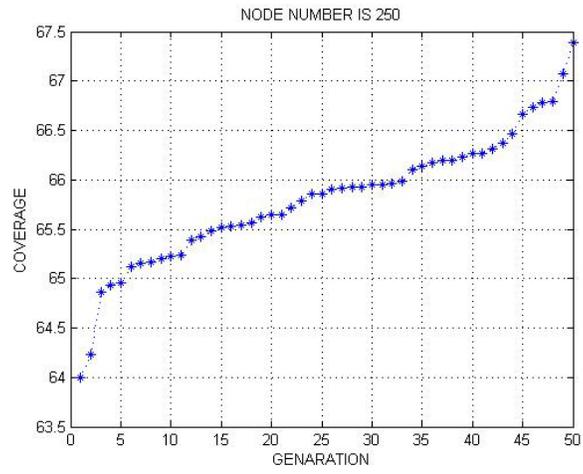

Figure 9

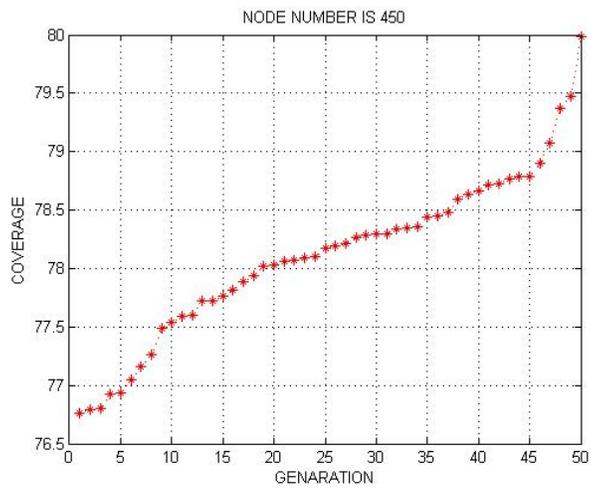

**Figure 10:**

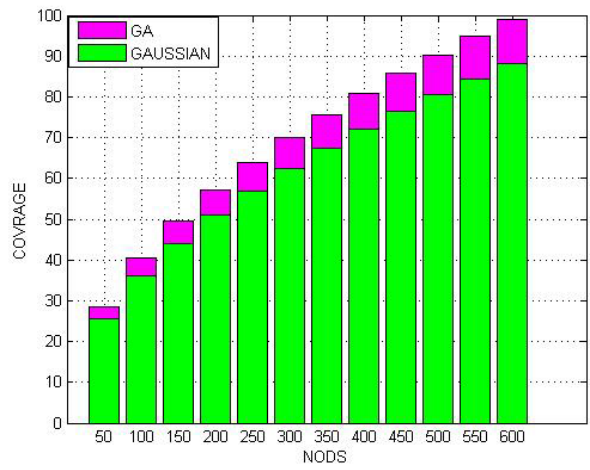

Figure 11

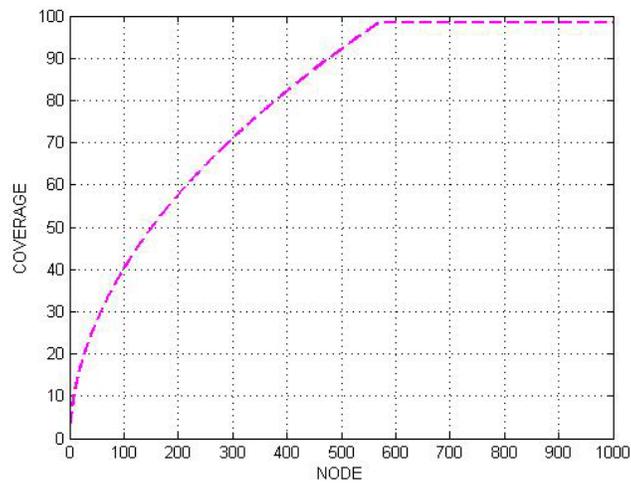